\documentclass[aps,prl,twocolumn,superscriptaddress,nofootinbib]{revtex4-1}

\usepackage{siunitx}
\usepackage{graphicx}
\usepackage[colorlinks=true,allcolors=black,pdfborder={0 0 0}]{hyperref}
\usepackage[all]{hypcap} 
\usepackage{booktabs} 
\AtBeginDocument{
	\heavyrulewidth=.08em
	\lightrulewidth=.05em
	\cmidrulewidth=.03em
	\belowrulesep=.65ex
	\belowbottomsep=0pt
	\aboverulesep=.4ex
	\abovetopsep=0pt
	\cmidrulesep=\doublerulesep
	\cmidrulekern=.5em
	\defaultaddspace=.5em
}

\begin{document}

	\title{Size-scaling limits of impulsive elastic energy release from a resilin-like elastomer}
	\author{Mark Ilton}
	\affiliation{Department of Polymer Science \& Engineering, University of Massachusetts Amherst, Amherst, MA, 01003}
	\author{S. M.  Cox}
	\affiliation{Department of Kinesiology, The Pennsylvania State University, University Park, PA, 16802}
	\author{Thijs Egelmeers}
	\affiliation{Department of Polymer Science \& Engineering, University of Massachusetts Amherst, Amherst, MA, 01003}
	\author{Gregory P. Sutton}
	\affiliation{School of Biological Sciences, University of Bristol, Bristol BS8 1UG, UK}
	\author{S. N. Patek}
	\affiliation{Department of Biology, Duke University, Durham, NC 27708}
	\author{Alfred J. Crosby}
	\email{crosby@mail.pse.umass.edu}
	\affiliation{Department of Polymer Science \& Engineering, University of Massachusetts Amherst, Amherst, MA, 01003}




\begin{abstract}
Elastically-driven motion has been used as a strategy to achieve high speeds in small organisms and engineered micro-robotic devices. We examine the size-scaling relations determining the limit of elastic energy release from elastomer bands with mechanical properties similar to the biological protein resilin. The maximum center-of-mass velocity of the elastomer bands was found to be size-scale independent, while smaller bands demonstrated larger accelerations and shorter durations of elastic energy release. Scaling relationships determined from these measurements are consistent with the performance of small organisms which utilize elastic elements to power motion. Engineered devices found in the literature do not follow the same size-scaling relationships, which suggests an opportunity for improved design of engineered  devices. 
\end{abstract}

\maketitle
\section{Introduction}

Many organisms  use impulsive, elastically-driven motion to exceed the power limitations of muscle ~\cite{James2007,Patek2011}. For example, mantis shrimp store elastic bending energy in the exoskeleton of their raptorial appendages, which upon release drives their appendages at velocities up to 30 m/s, allowing them to crush shells or spear prey~\cite{Patek2004,Zack2009,Rosario2015,McHenry2016}. Although large organisms make some use of elastic structures (e.g. tendon), elastic energy storage and release can be crucial for small organisms (typically $\!<\!10\,\mathrm{cm}$ in length)  to achieve rapid movement~\cite{Bennet-Clark1977,James2007}. These small organisms - such as mantis-shrimp, trap-jaw ants, locusts and fleas - use a latch to separate the phase of elastic energy storage (via muscle contractions) from that of energy release~\cite{Ilton2018}. Disentangling the rate of muscle contraction from energy release allows these organisms to achieve astonishing kinematic performance (high velocities, large accelerations, and short durations of movement), and perhaps most remarkably, to perform these motions in a repeatable manner sustained by their metabolic processes. 

Organisms that store and release elastic energy have served as inspiration for recent robotics research~\cite{Kovac2008,Debray2011,Koh2013,Zhao2013,Kim2014a,Koh2015,Zaitsev2015,Haldane2016}. Several research groups have taken a biomimetic or bioinspired approach in an attempt to match (or exceed) biological performance using engineered devices. This approach has led to new techniques for robotic manipulation~\cite{Debray2011,Rollinson2013,Kim2014a,Haldane2016}, the ability to move robots on difficult terrain~\cite{Kovac2008,Koh2015,Zaitsev2015}, and has been used to test scientific hypotheses about locomotion~\cite{Koh2013,Kim2014a,Koh2015}. However, these engineered devices are typically larger than biological organisms, and the fastest organisms have a greater kinematic performance than currently achievable by small robots using elastic elements to perform repeatable motions~\cite{Ilton2018}.

In addition to an elastic element (i.e. spring) there are three other major components of an elastically-driven system: (i) a motor (in many animals, muscle) that generates sufficient work to load the elastic element, (ii) an energy-efficient latch to store and release the elastic element without significant dissipation, and (iii) a load mass that is moved by the elastic element and that is not actively involved in elastic energy release. But since these systems drive motion through elastic recoil, the kinematic performance in these systems depends on the properties of an elastic element. Although springs are often assumed to be ideal, the materials properties and geometry of a spring can constrain its kinematics~\cite{Ilton2018}. 
 
 In this work, we address the gap in performance between biological and synthetic systems by examining the role of size-scale and materials properties for elastic energy storage and release. To determine the limits of elastic energy release due to only spring properties, we take a reductionist approach by examining the dynamics of a freely-retracting spring in isolation - externalizing the motor and latch. This externalization decouples the motor and latch from the fast movement of the spring, which is similar to the way some fast elastically-driven organisms operate~\cite{Cox2014}. Here we take this isolation one step further by measuring the dynamics of a spring that carries no additional load mass. From an initially uniform uniaxial extension, we release long thin bands of polyurethane elastomer. This polyurethane has similar mechanical properties to resilin, an elastomeric protein found in some arthropods and important for elastically-driven motion in locusts~\cite{Burrows2012,Burrows2016}. Resilin is a material with high resilience (resilience is a measure of energy recovery, and is defined by the ratio of energy recovered upon unloading divided by the energy expended during loading a material) with $r>$90\% resilience measured for both natural and recombinant resilin~\cite{Weis-Fogh1961,Jensen1962,King2010,Gosline2002,Elvin2005}. We measure the free retraction of a resilin-like polyurethane elastomer, and building upon recent work~\cite{Vermorel2007,Bogoslovov2007,Niemczura2011,Tunnicliffe2015,Yoon2017}, track the full displacement field of the material. The displacement field is used to obtain the center-of-mass motion of the band, which allows for a functional determination of the scaling relations that define the limits of impulsive elastic performance. We focus on the size-scale and materials properties of a spring and how these factors affect its elastically-driven performance by examining three key parameters often used to assess kinematic performance in biology and micro-robotics~\cite{James2007,Patek2011,Patek2004,Kovac2008,Rogers2016,Burrows2006,Gerratt2013,Irschick2003}: maximum center-of-mass velocity ($v_\mathrm{max}$), maximum center-of-mass acceleration ($a_\mathrm{max}$), and duration of elastic energy release ($\Delta t$). Utilizing this experimental approach we ask two guiding questions: Does kinematic performance depend on the size of an elastic element?  How does the kinematic performance of elastically-driven biological organisms and engineered devices compare to the isolated recoil of a resilin-like elastomer? 

Expected scaling relations for the center-of-mass kinematic performance of a recoiling elastomer band can be rationalized based on physical principles. First, the center-of-mass acceleration of the band is given by the ratio of the net force acting on the band divided by its mass. Just after the release of the band from one end, if the only external force acting on the band is from the clamp at the other (fixed) end, then center-of-mass acceleration is

\[
a_\mathrm{max} = \frac{\sigma_\mathrm{in}}{\rho L_0},
\]

where $\sigma_\mathrm{in}$ is the initial stress from which the band is released, while $L_0$ and $\rho$ are the equilibrium length of the band and its density, respectively. To separate the role of materials and loading strain, we can rewrite this equation  as

\begin{equation}
\label{eq:a}
a_\mathrm{max} = \frac{c_\mathrm{sec}^2 \epsilon_\mathrm{in}}{L_0},
\end{equation}

with the secant elastic wavespeed ($c_\mathrm{sec}$) from an initial strain ($\epsilon_\mathrm{in}$) defined as

\begin{equation}
\label{eq:csec}
c_\mathrm{sec} = \sqrt{\frac{\sigma_\mathrm{in}}{\rho \epsilon_\mathrm{in}}}.
\end{equation}

During the unloading of a uniform, long thin strip of  elastic material stretched to an initial strain of $\epsilon_\mathrm{in}$, the center-of-mass travels a displacement $\epsilon_\mathrm{in} L_0/2$. Using this displacement and assuming a constant center-of-mass acceleration given by Eq.~(\ref{eq:a}), leads to the duration of elastic energy release
\begin{equation}
\label{eq:t}
\Delta t = \frac{L_0}{c_\mathrm{sec}}.
\end{equation}

Finally, with those same assumptions, the maximum center-of-mass velocity is determined by the product of acceleration and duration, $v_\mathrm{max} = a_\mathrm{max} \Delta t$, yielding an expression consistent with the maximum velocity found in previous work for a linear elastic material~\cite{Tunnicliffe2015}
\begin{equation}
\label{eq:v}
v_\mathrm{max} = c_\mathrm{sec} \epsilon_\mathrm{in}.
\end{equation}

\section{Materials and Methods}

\begin{figure}[htp]
	\centering
	\includegraphics[width=75mm]{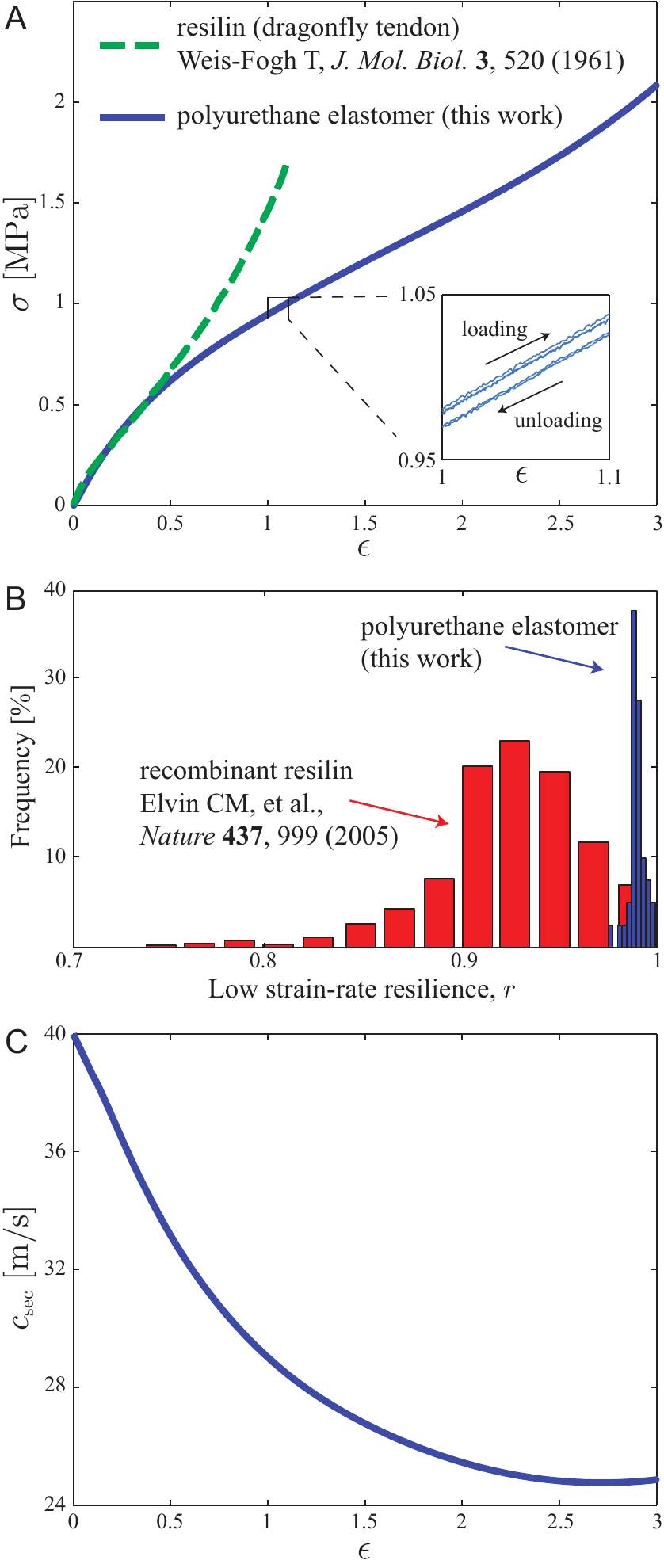}
	\caption{\label{fig:1}\textbf{Low strain-rate characterization of a resilin-like polyurethane elastomer.} \textbf{A} The polyurethane elastomer used in this study is similar to resilin in its mechanical properties at low strain-rate ($\dot{\epsilon}<0.01$). Its stress-strain response has a similar slope (modulus) to resilin (from ref.~\cite{Weis-Fogh1961}) at small strain in uniaxial extension. The small difference in stress between loading and unloading the polyurethane (inset) can be quantified by the material's resilience. \textbf{B} The polyurethane elastomer has a high resilience ($r>0.97$) for all samples measured in this study, up to $\epsilon=3$. Its resilience is similar to natural resilin ($r=0.96-0.97$)~\cite{Jensen1962}, and slightly higher than the recombinant resilin from ref.~\cite{Elvin2005}. \textbf{C} The secant wavespeed ($c_\mathrm{sec}$) depends on strain for the polyurethane elastomer, as determined from the stress-strain response and Eq.~(\ref{eq:csec}). As will be shown, $c_\mathrm{sec}$ is a characteristic velocity that governs the recoil dynamics of the elastomer. }
\end{figure}

To experimentally verify these scaling relations, a commercially available pre-fabricated polyurethane elastomer sheet (McMaster-Carr, 8716K61, durometer 40A, $1.6\,\mathrm{mm}$ thick) was sectioned into long, thin bands using a razor blade. For the narrowest bands (width,  $w_0\!<\!2\,\mathrm{mm}$), a laser cutter (Universal Laser Systems) was first used to create shallow grooves to guide the razor blade, reducing variation in the band width. Samples were cut to ensure a uniaxial geometry ($L_0\!>>\!w_0$), with $1.6 \,\mathrm{mm}\! \leq\! w_0 \!\leq\! 27 \,\mathrm{mm}$ and $17\, \mathrm{mm}\! \leq\! L_0\! \leq\! 267\, \mathrm{mm}$. The mechanical properties of the material were characterized by performing cyclic loading/unloading of the bands at low strain-rate ($\dot{\epsilon}< 0.01\, \mathrm{s}^{-1}$), using a tensile testing apparatus (Instron 5564). The polyurethane elastomer has similar mechanical response to resilin (Fig.~\ref{fig:1}A-B) at low strain-rates, and a resilience $r>97\%$ at up to 300\% strain. Beyond 300\% strain, the material would typically fail due to stress concentrations at the clamped ends of the band. While resilin can strain up to 300\% reliably~\cite{Weis-Fogh1961}, it is not generally observed to stretch this much in vivo. The highest suggested in vivo strain for a recoiling insect spring is in the flea pleural arch, where the resilinous portion is hypothesized to strain 100\%~\cite{Bennet-Clark1967} - thus our experimental elastomer strain covers the whole range of the strains seen in vivo. Over the full range of the polyurethane elastomer, the secant wavespeed of the polyurethane depends on strain and varies between $\sim 24 - 40\,\mathrm{m/s}$ (Fig.~\ref{fig:1}C), as calculated from Eq.~\ref{eq:csec} using the stress-strain relationship in Fig.~\ref{fig:1}A and the density of the material (average density of all samples $\rho = 1125 \,\mathrm{kg/m}^3$, with the mass of each sample measured using an analytical balance).

Free retraction measurements were performed by initially loading a band clamped between two pneumatic grips to a given initial strain ($\epsilon_\mathrm{in}$) using the tensile testing apparatus, and then releasing one of the grips (Fig.~\ref{fig:2}A). Upon release, the band rapidly contracts, and the motion was recorded using a high speed camera (Photron Fastcam SA3, frame rate 20-75 kfps). A macro zoom lens (Nikon AF Nikkor 24-85mm) was used to maximize the image of the band to cover the full 1024 pixel CCD of the camera along the direction of motion ($x$-axis), giving a pixel resolution of $33-\SI{420}{\micro \meter}$ depending on the band length and initial strain. Markings, which had been placed along the band (Sharpie\textsuperscript{\textregistered{}}  marker, metallic silver), were then digitized from the high speed videography using a custom MATLAB script to determine the position ($x$) of each point of the band as a function of time ($t$) (Fig.~\ref{fig:2}B).  To generate velocity, acceleration, and higher order derivatives of the position with respect to time, the digitized position data was fit to free knot splines~\cite{Schwetlick1995,Schutze1997}. Combining the motion of each section of the band, the center-of-mass kinematics were then deduced, allowing for the determination of $v_\mathrm{max}$ and $a_\mathrm{max}$. The duration was defined as the time between the onset of the propagating elastic wave (determined by a minimum onset threshold of jerk) until the kinetic energy of the band reached its maximum (which occurs at $v=v_\mathrm{max}$). 

\section{Results}
\begin{figure}[htp]
	
	\centering
	\includegraphics[width=0.96\columnwidth]{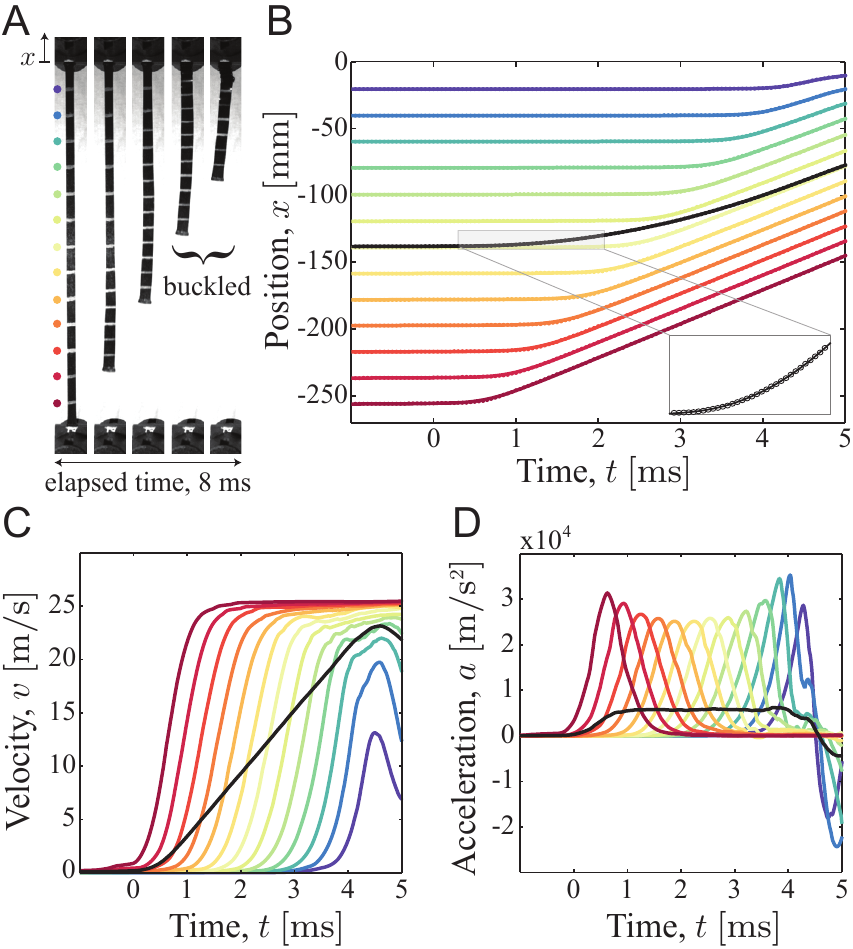}
	\caption{\label{fig:2}\textbf{The center-of-mass kinematic performance (velocity, acceleration, and duration) is measured for a retracting elastomer band.} \textbf{A} Five images of a retracting elastomer band ($L_0\!=\!140\,\mathrm{mm}$, $w_0\!=\!8.5\,\mathrm{mm}$) from a high speed image sequence. To visualize the motion of the band, silver markings are placed along the band and on the clamps at the top and bottom of the image (colored points to the left of the images were added in post-processing to uniquely label the points of the band, and correspond to the colors used in B-D). The last two images in the sequence show the band undergoing compressive buckling, and occur after the center-of-mass has reached its maximum velocity. \textbf{B} After the bottom clamp releases, motion propagates up through the band in a spatially non-uniform release of strain energy. The center-of-mass motion (black points) is determined from a weighted average of the individual segments of the band (colored points). The inset shows a zoom in of the center-of-mass position along with a free knot spline fit (solid, black curve) in close agreement with the data. The maximum  difference between the data and spline fit is on the order of a single pixel ($\sim 0.2$\% of the total displacement of the center-of-mass).  $t\!=\!0$ is set by the propagation of the elastic wave unloading, and determined by a minimum threshold in the derivate of the acceleration (jerk). \textbf{C,D} Derivatives of the free knot splines give the velocity (C), and acceleration (D) of each segment of the band (colored curves), along with the center-of-mass (black curve). From the center-of-mass velocity and acceleration, the kinematic performance is determined (here $v_\mathrm{max} = 23\,\mathrm{m/s}$, $a_\mathrm{max} = 6.2\times 10^3\,\mathrm{m/s}^2$, $\Delta t = 4.6\,\mathrm{ms}$).}
\end{figure}

\begin{figure*}[t!]
	\centering
	\includegraphics[width=180mm]{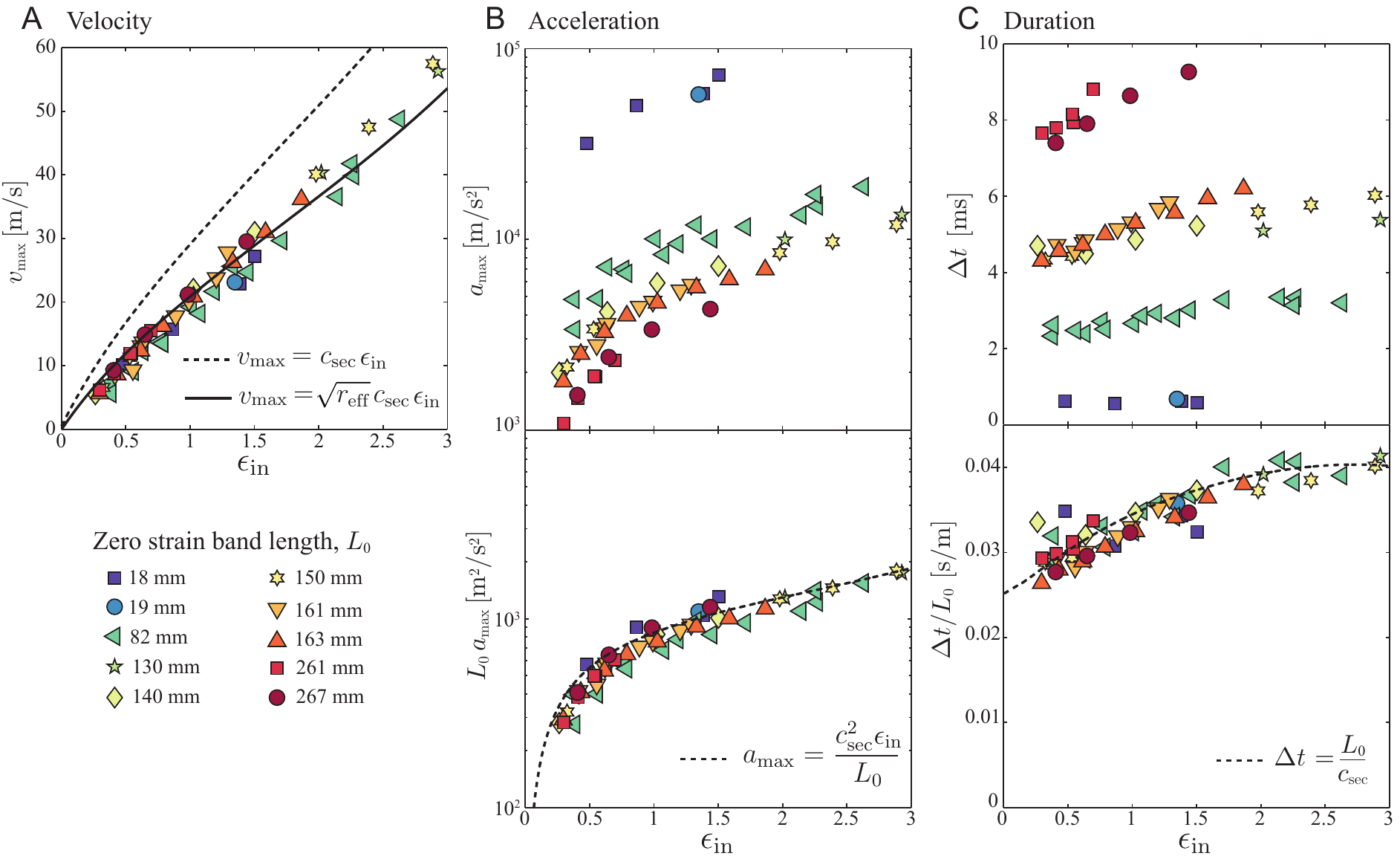}
	\caption{\textbf{Scaling relations from Eqs. (\ref{eq:a}-\ref{eq:v}) are tested by recoil experiments.} \textbf{A} The maximum velocity from 57 recoil measurements increases with the initial strain before release ($\epsilon_\mathrm{in}$), and is independent of the band length. \textbf{B} The maximum acceleration increases with initial strain, and varies inversely with the band length (bottom panel shows data collapsed in terms of $L_0 a_\mathrm{max}$). \textbf{C} Recoil duration has a weak (convex downward) dependence on initial strain, and varies inversely with band length (bottom panel shows collapsed data). The dashed curves are the predicted scaling relations from Eqs. (\ref{eq:a}-\ref{eq:v}), while the solid curve in panel A is from Eq.~(\ref{eq:v_eff}). \label{fig:3} }
\end{figure*}

The kinematic performance of 13 different bands with varying geometry (varying $L_0$ and $w_0$) was measured (Fig.~\ref{fig:2} shows an example measurement) as a function of the strain energy loaded into the band (between 1-8 values of $\epsilon_\mathrm{in}$ for each band, for a total of 57 unique measurements). The center-of-mass kinematic performance does not depend on $w_0$ for the uniaxial geometry used in these experiments. The maximum center-of-mass velocity, acceleration, and duration all  increase with increasing initial strain (Fig.~\ref{fig:3}A-C). The center-of-mass velocity is independent of the band length  (Fig.~\ref{fig:3}A), however, the maximum center-of-mass acceleration and duration both depend on band length (Fig.~\ref{fig:3}B-C); the acceleration is inversely proportional to band length (Fig.~\ref{fig:3}B, bottom panel) and the duration scales with band length (Fig.~\ref{fig:3}C, bottom panel), as demonstrated by the data collapse after appropriately normalizing $a_\mathrm{max}$ and $\Delta t$ with $L_0$.

The scaling relations predicted by Eqs. (\ref{eq:a}-\ref{eq:v}) are comparable to the observed recoil kinematics (dashed curves in Fig.~\ref{fig:3}A-C), using $c_\mathrm{sec}$ measured from the tensile test (Fig.~\ref{fig:1}C). The scalings agree with the data for acceleration and duration with no free parameters (Fig.~\ref{fig:3}B-C). However, the scaling relationship for velocity systematically exceeds the observed recoil velocity (Fig.~\ref{fig:3}A). 

To understand the systematic difference in predicted and measured recoil velocity, it is helpful to examine the predicted velocity scaling of  Eq.~(\ref{eq:v}) in the context of the kinematic data in Fig.~\ref{fig:2}D. The equation assumes a constant acceleration over the entire duration recoil. Although this is a reasonable approximation, the measured duration also includes the ramp-up time to reach $a_\mathrm{max}$ ($\sim 1\,\mathrm{ms}$ in Fig.~\ref{fig:1}D)  and the ramp-down to zero acceleration (also $\sim 1\,\mathrm{ms}$ in Fig.~\ref{fig:1}D). During this ramp-up and ramp-down period the acceleration is less than $a_\mathrm{max}$, which leads to a breakdown in the predicted scaling of Eq.~(\ref{eq:v}). Factors that could affect the ramp-up/ramp-down time include frictional losses from interaction of the band with the pneumatic clamp~\cite{Tunnicliffe2015}, inertia of elastomer material inside the clamp, dispersion of the elastic wave due to losses within the material or to the environment~\cite{Vermorel2007}, and residual strain left in the band at the point of buckling ~\cite{Bogoslovov2007}. These losses depend on both  material properties of the band and external factors. Since these factors are challenging to accurately model, as a first approximation we assume these losses are constant for all the bands measured, and introduce an effective resilience of the recoiling elastomer $r_\mathrm{eff}$ through the scaling 

\begin{equation}
\label{eq:v_eff}
v_\mathrm{max} = \sqrt{r_\mathrm{eff}}c_\mathrm{sec}\epsilon_\mathrm{in}.
\end{equation}

This effective resilience accounts for both energy loss within the band and external dissipation (such as friction of the clamp), and $r_\mathrm{eff}$ is defined by the ratio of output kinetic energy to input elastic energy (i.e. $r_\mathrm{eff} = \rho v_\mathrm{max}^2/2 u_\mathrm{in}$, where $u_\mathrm{in}$ is the stored elastic energy density). The energy loss seems to primarily occur during the ramp-up and ramp-down periods of the recoil, which accounts for why $a_\mathrm{max}$ and $\Delta t$ do not depend on $r_\mathrm{eff}$. Using $r_\mathrm{eff}$ as a free parameter to fit the measured recoil velocity (Fig.~\ref{fig:3}A, solid curve), results in $r_\mathrm{eff} = 0.5\,\pm 0.1$. This effective resilience is significantly lower than that measured at low strain-rate (recall $r>0.97$ from Fig.~\ref{fig:1}B).  

\section{Discussion}

\begin{figure*}[t]
	\centering
	\includegraphics[width=1\textwidth]{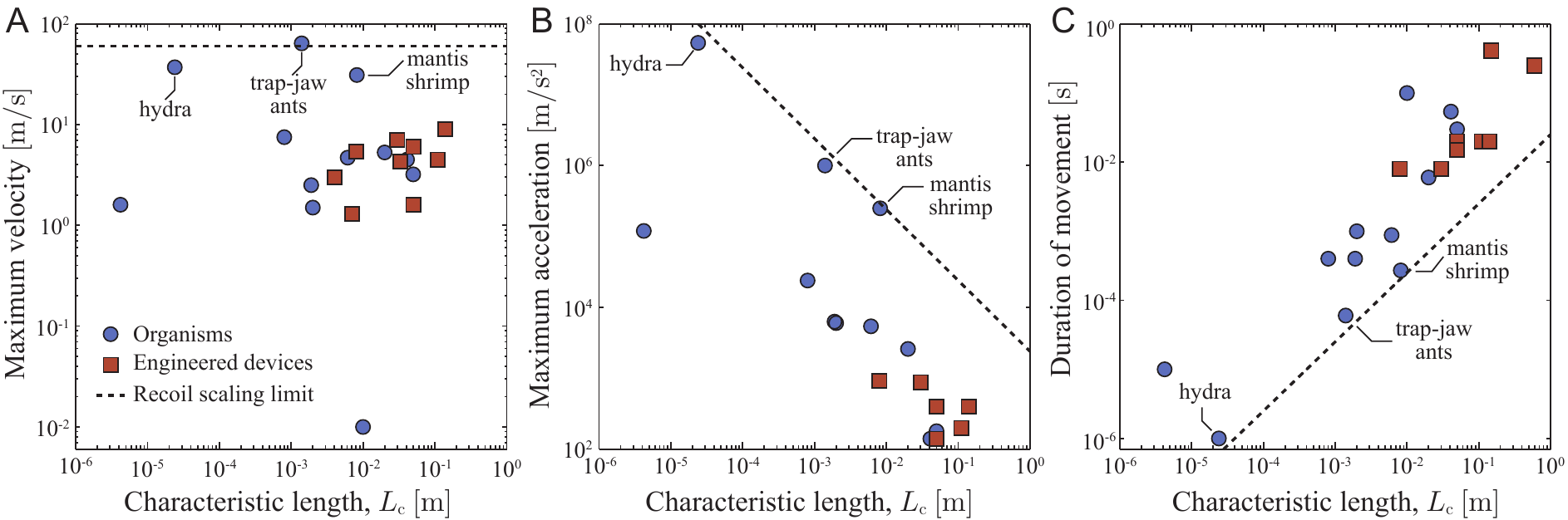}
	\caption{\textbf{Scaling relations from recoil experiments have a similar size scaling as organisms which use elastically-driven motion and show similar performance limits.} The kinematic performance for organisms and engineered devices (from refs.~\cite{Zaitsev2015a,Niiyama2007,Kim2014a,Kajita2003,Kovac2008,Koh2015,Churaman2012,Forterre2005,Burrows2006,Vincent2011,Zhao2013,Koh2013,Debray2011,Gerratt2013,Bennet-Clark1975,Anderson2016,Anderson2012,James2008,McHenry2016,Burrows2012a,Patek2006,Whitaker2007,Edwards2005,Pringle2005,Koch1998,Nuchter2006}, summarized in Table 1 of ref.~\cite{Ilton2018}) are compared with the limits of the recoil measurements (this work) in terms of \textbf{A} velocity, \textbf{B} acceleration, and \textbf{C} duration. The dashed line is taken using the scaling relations found in Fig.~\ref{fig:3} (using $\epsilon_\mathrm{in}\!=\!3$).\label{fig:4}}
\end{figure*}

Armed with scaling relations that agree with the observed recoil kinematics, we now return to answering our first guiding question: Does kinematic performance depend on the size of an elastic element? The size-scaling limits of the resilin-like polyurethane elastomer for repeatable, elastic energy release (Fig.~\ref{fig:4}, dashed lines) are determined by setting the initial strain to $\epsilon_\mathrm{in}=3$ in the scaling relations from Fig.~\ref{fig:3} (recall for $\epsilon_\mathrm{in}> 3$ failure of the polyurethane was often observed). The maximum velocity of the polyurethane elastomer recoil is size-scale independent (Fig.~\ref{fig:4}A), while the maximum acceleration and duration of movement depend on size (Fig.~\ref{fig:4}B-C). The dashed lines in Fig.~\ref{fig:4} represent the kinematic performance of this particular material choice of polyurethane elastomer, under a specific loading geometry (uniaxial extension), and driving zero added load mass. In the next two paragraphs we justify two specific claims about the recoil scaling limits shown in Fig.~\ref{fig:4}: (1) the overall scaling of kinematic performance with size does not depend on the specific choice material, geometry, or load mass, and (2) the dashed lines in Fig.~\ref{fig:4} are an approximate upper bound for the particular material choice used in this study, independent of geometry and load mass. 

First, the size-scaling of kinematic performance (summarized in the first column of Table~\ref{tab:power}) should be independent of the specific choice of materials, geometry, and load mass. Changing the elastic material would alter the pre-factors in the scaling limits through changing $c_\mathrm{sec}$, failure properties, and resilience, without altering the fundamental trade-offs with size-scale~\cite{Ashby2011materials}. A different geometry (e.g. using a cantilevered beam as a spring) or adding load mass to the system would alter the absolute kinematic performance of the system. However, if the relative size of elements all change with system size, then changing geometry or mass simply introduces a lengthscale-independent pre-factor to the scaling relations. As a specific example, for a cantilevered beam driving a heavy load mass the scaling relations shown in Fig.~\ref{fig:3} still hold, but with added coefficients that depend on two dimensionless parameters: the aspect ratio of the beam (length to thickness), and the ratio of the spring mass to load mass (see Supplementary Information). Since these are independent of size-scale when relative size proportions are held constant, the scaling of kinematic performance with characteristic length shown in Fig.~\ref{fig:4} are robust descriptions of the size-scale dependence of elastically-driven motion. 

\begin{table}[b]
	\centering
	\caption{Dependence of velocity, acceleration, and duration on the characteristic lengthscale ($L_c$) for recoil measurements along with two parameter power law fits to the organisms and engineered devices in Fig.~\ref{fig:4}. Here we report the power law exponent $\alpha$, obtained by fitting to $AL_c^\alpha$, where both $A$ and $\alpha$ are adjustable fitting parameters.\label{tab:power}}
	
	\begin{tabular}{p{2.0cm} p{1.5cm} p{1.6cm} p{1.5cm}}
		\toprule
		& Recoil & Organisms & Devices \\ \midrule
		Velocity & $\sim L_c^{0}$  &  $\sim L_c^{-0.1}$ &  $\sim L_c^{0.2}$\\ 
		Acceleration &  $\sim L_c^{-1}$ & $\sim L_c^{ -0.9}$ &  $\sim L_c^{-0.5}$  \\  
		Duration &  $\sim L_c^{1}$ &  $\sim L_c^{1.1}$ &  $\sim L_c^{0.9}$   \\
		\bottomrule
	\end{tabular}
\end{table}

The second claim above --- that for the specific polyurethane used in this study the dashed lines in Fig.~\ref{fig:4} are an approximate upper bound to elastically-driven performance ---  is also related to the geometry and load mass. In both of these cases, changing geometry or adding load mass, the net effect is a decrease in the system's kinematic performance and does not change the scaling argument in Fig.~\ref{fig:4}. Intuitively, adding load mass to the system would decrease the kinematic performance compared to the unloaded elastomer bands used here. The uniaxial geometry used in this work ensures a nearly uniform strain energy density in the material. Other geometries (such as bending) result in a non-uniform strain energy density, and material failure will likely occur at a lower average strain energy density than for uniaxial extension (see Supplementary Information). And although geometries which introduce a mechanical advantage in the system through a lever arm can amplify displacement, they also increase inertial load. As a result, a longer lever arm does not improve performance of the three kinematic parameters in Fig.~\ref{fig:4}. Therefore, changing geometry or load mass would shift the polyurethane scaling to lower performance (lowering the intercepts in the plots of Fig.~\ref{fig:4}), without altering the size-scaling relationship (the slopes in Fig.~\ref{fig:4}).

Putting these results in a larger context, we return to our second guiding question:  How does the kinematic performance of elastically-driven biological organisms and engineered devices compare to the isolated recoil of a resilin-like elastomer? We interpret our results by comparing the size-scale dependence of the kinematic performance of the model elastomer with the performance of organisms and engineered devices that incorporate elastic elements (Fig.~\ref{fig:4}).  The limits of kinematic performance for the polyurethane elastomer shows a similar size-scaling to elastically-driven organisms, which is in contrast to the engineered devices (Table~\ref{tab:power}). Specifically, the maximum acceleration scales inversely with characteristic lengthscale for both elastic recoil measurements and organisms, yet the maximum acceleration of current engineered devices depends more weakly on size-scale. We are cautious in the interpretation of this result as each organism or engineered device in this dataset represents a unique embodiment of materials properties and geometry of elastic energy release, and the engineered devices span a narrower range of lengthscales than the organisms. However, the connection between size-scale dependence of the recoil performance and elastically-driven organisms suggests a possible universality to the size-scaling limits of elastic energy release.

Another notable feature that emerges from Fig.~\ref{fig:4} is the ability for examples from biology to match the performance of the synthetic elastomer system. The scaling limits of kinematic performance for the elastomer recoil is similar to the performance of hydra, trap-jaw ants, and mantis shrimp. This is impressive for three reasons. First, compared to our isolated polyurethane elastomer, we would expect a diminished performance for organisms because they have load mass that does not contribute to elastic energy storage. For example, in the raptorial appendage of mantis shrimp, the two regions that move furthest (the propodus and dactyl) do not store significant elastic energy~\cite{Rosario2015}, and the added mass of these regions slows the release of elastic energy. Second, dissipation is likely much more significant at the lengthscales of these organisms~\cite{Bennet-Clark1977}, and remarkably, both the hydra and mantis shrimp achieve their kinematic performance under water in a viscous environment. Finally, since performance of organisms in the lab is often inferior to that in nature~\cite{Astley2013,Irschick2003}, the kinematic performance of these organisms could potentially be higher in a natural setting. The remarkable performance of hydra, trap-jaw ants, and mantis shrimp despite these hindering factors, suggests that the materials properties of the biological springs are likely critical to their kinematics. While resilin is often discussed as an energy store (going back to refs. \cite{Weis-Fogh1961} and \cite{Bennet-Clark1967}), many arthropods also use the much harder chitin as a primary material to store energy, as is the case for chitinous springs in locusts~\cite{Burrows2012,Wan2016}, froghoppers~\cite{Burrows2008,Siwanowicz2017}, planthoppers~\cite{Siwanowicz2017}, mantis shrimp~\cite{Tadayon2015}, and trap-jaw ants~\cite{Larabee2017}. Chitin, having an elastic modulus orders of magnitude larger than resilin~\cite{Vincent2004a}, may account for the ability of arthropod systems to surpass the maximums observed in our experiments which use a resilin-like elastomer as the primary energy store. This difference in modulus is consistent with observed elastic mechanisms in the highest performing organisms, including the chitinous exoskeletal elastic materials in mantis shrimp~\cite{Patek2004} and trap-jaw ants~\cite{Gronenberg1996}, along with mini-collagen fibrils in hydra~\cite{Nuchter2006}.

Material properties of elastic elements have been shown previously to play an important role in elastic energy storage and release in synthetic systems. Work on engineered devices has noted the importance of using spring materials with a high elastic energy density capacity, such as elastomers~\cite{Lee1990,Ashby2011materials}. Even though metals typically have a significantly higher elastic wavespeed, the large strain to failure of elastomers allows them to reach velocities that are often greater than that of metal springs~\cite{Maier1957}. However, typical elastomers dissipate a significant fraction of the stored elastic energy (low resilience), so one might expect that elastomeric materials with high resilience, such as elastin or resilin found in some organisms~\cite{King2010,Gosline2002,Elvin2005}, would serve as ideal candidates for the quick release of elastic energy. Understanding the trade-offs between resilience, elastic wavespeed, and maximum strain in biological materials employed by organisms undergoing elastically-driven motion could provide insight into the ultimate limits of elastic energy release. Recent evidence indicates that similar trade-offs persist in biological systems. While resilin and elastin are highly resilient materials, their capacity for elastic energy storage is low. This suggests that the coupling of resilient and stiff materials commonly found in biological systems may offset these inherent trade-offs~\cite{Burrows2008}. The weak size-scale dependence of the engineered devices  (Table~\ref{tab:power}) and their diminished performance compared to biological organisms demonstrates that there are opportunities for improved design. For small-scale devices, performance enhancements could be developed from a bioinspired approach, utilizing composite elastic materials with both resilient and stiff components. Depending on the desired function of the device, this work suggests some advantage to engineering devices at smaller size-scales to maximize performance (e.g. maximizing acceleration or minimizing duration might be a goal). A deeper understanding of spring performance in the context of an on-board motor, latch, and load mass could further reveal important design principles currently limiting engineering design.

The impact of size-scale on kinematic performance is complicated by the choice of using either absolute performance, or scaling the performance relative to body size (relative performance). Biologists examine kinematics of organisms both in an absolute sense (a cheetah runs more quickly than an ant) and in a relative sense (relative to body size, some ants are faster than cheetahs). Relative performance of running (body lengths per second) and jumping (jump height per body length) have been used to characterize both biological organisms~\cite{Avery1989,Spagna2012,Burrows2006,Rosario2016} and engineered devices~\cite{Kovac2008,Saranli2001,Gerratt2013,Gao2012}. In biology, relative size has been used to standardize for size differences between animals in the same species~\cite{Avery1989} or across several species~\cite{Spagna2012}, and it has been suggested that relative performance is more ecologically relevant as it correlates well with the ability to evade predators~\cite{VanDamme1999,Bergmann2009}. Relative performance can also be used to normalize for drag effects, which become significant at small size-scales~\cite{Bennet-Clark1977}. However, in contrast to the prevalent use of relative performance, we find that \emph{absolute} velocity is a size-scale independent quantification of elastic performance for the lengthscales probed in the current work. Higher \emph{relative} velocities (along with higher accelerations or shorter durations) can be achieved simply by reducing the size of an elastic element. Therefore, comparing the performance of systems that are orders of magnitude different in size-scale requires caution.

In summary, we have measured the kinematic performance of elastic energy release for an elastomer with similar mechanical properties to the protein resilin. In agreement with expected scaling relations, the maximum center-of-mass velocity of a freely retracting band is independent of length, and depends only on the initial strain at which the band was released and the elastic wavespeed of the material. The maximum center-of-mass acceleration and duration of elastic energy release were found to depend on the length of the elastomer band, with an improved performance at smaller size-scales.  Previously reported measurements of kinematic performance in elastically-driven organisms show similar size-scaling limits to the elastomer studied here, whereas the acceleration of engineered micro-robotic devices varies more weakly with size-scale. The current results, which probe the upper bound of elastically-driven kinematics of a resilin-like material, show a similar performance to some of the fastest biological systems. Future work which seeks to delineate the role of elastic wavespeed, maximum strain, and resilience in elastic biological systems could lead to a foundational understanding for improved engineering design. Specifically, the mechanical properties of resilin, chitin, and resilin/chitin composites would be of great importance to compare to engineered systems.

\subsection*{Data accessibility.} All relevant data are included in the manuscript. This
article has no additional data.
\subsection*{Authors' contributions.} M.I. S.M.C., A.J.C. designed research; M.I., T.E. performed experiments; M.I., S.M.C., G.P.S., S.N.P., A.J.C. wrote the paper.

\subsection*{Competing interests.} The authors declare no conflict of interest.

\subsection*{Funding.}This material is based upon work supported by the U. S. Army Research Laboratory and the U. S. Army Research Office under contract/grant number W911NF-15-1-0358.

\section*{Acknowledgments.}
The authors thank Professor Ryan Hayward and Tetsu Ouchi for help with high speed imaging.

\end{document}